\documentclass[prl,reprint,amsmath,amssymb,amsfonts]{revtex4-1}

\usepackage[draft]{changes}
\usepackage{color}
\usepackage{nccmath}
\usepackage{mathrsfs}
\usepackage{esint}
\usepackage{bm}
\usepackage{cancel}
\usepackage[normalem]{ulem}
\usepackage{graphicx}
\usepackage[caption=false]{subfig}
\usepackage[unicode=true,pdfusetitle,bookmarks=true,
bookmarksnumbered=false,bookmarksopen=false,breaklinks=false,
pdfborder={0 0 0},backref=false,colorlinks=true,citecolor=red]
{hyperref}
\usepackage{cleveref}
\providecommand\eqref[1]{\ref{eq:#1}}
\renewcommand\b[1]{{\bf  #1}}
\renewcommand\vec[1]{\boldsymbol{#1}}
\renewcommand\phi{\varphi}

\newcommand\del{\nabla}
\newcommand\dd{\mathrm{d}}

\newcommand{\KR}{K_\mathrm{eq}^R}
\newcommand{\Eh}{\mathsf{Eh}}
\newcommand{\Pe}{\mathsf{Pe}}

\allowdisplaybreaks[3]

\graphicspath{{figures/}}

\begin{document}

\title{Mechanical activity enables patterning and  discrimination at the immune synapse}
\author{Tony Wong$^1$}
\author{Tom Chou$^{1,2}$}
\author{Suraj Shankar$^3$}
\email[]{surajsh@umich.edu}
\author{Shenshen Wang$^4$}
\email[]{shenshen@physics.ucla.edu}
\affiliation{$^1$Department of Mathematics, UCLA, Los Angeles, CA, 90095-1555, USA}
\affiliation{$^2$Department of Computational Medicine, UCLA, Los Angeles, CA, 90095-1766, USA}
\affiliation{$^3$Department of Physics, University of Michigan, Ann Arbor, MI 48109, USA}
\affiliation{$^4$Department of Physics and Astronomy, University of California, Los Angeles, Los Angeles, CA 90095, USA}

%
%
%
%
%
%
%
%
\date{\today}

\begin{abstract}
    Immune cells recognize and discriminate antigens through immunological synapses -- dynamic intercellular junctions exhibiting highly organized receptor-ligand patterns. While much work has focused on molecular 
    kinetics and passive mechanisms of pattern formation, the role of active mechanical control in patterning and discrimination remains underexplored. We develop a minimal continuum model coupling receptor binding kinetics, membrane
    deformation, and cytoskeletal forces, with elastohydrodynamic flow in the synaptic cleft. Numerical simulations and scaling analysis reveal that contractile cortical flows arrest coarsening and stabilize long-lived multifocal clusters, whereas active pulling accelerates cluster dissolution and elevates background receptor binding. Nonequilibrium mechanical forces enable adaptive control over the speed, sensitivity, and dynamic range of affinity discrimination in a pattern-dependent manner. 
    Our results highlight how immune cells exploit cytoskeletal remodeling to robustly regulate antigen recognition through synaptic patterning.
\end{abstract}

\maketitle

\noindent
\textit{Introduction.} Cells sense dynamic environments, process information, and make decisions; these computational capabilities are critical to immune function. T and B lymphocytes of the adaptive immune system acquire information from antigen-presenting cells (APCs) by forming an immunological synapse~\cite{grakoui1999immunological,dustin2012receptor,GROVES2011}. This intercellular junction contains dynamic patterns of antigen-receptor complexes and adhesion molecules~\cite{dustin2010understanding}.

\begin{figure}[t]
    \centering
    \includegraphics[width=0.45\textwidth]{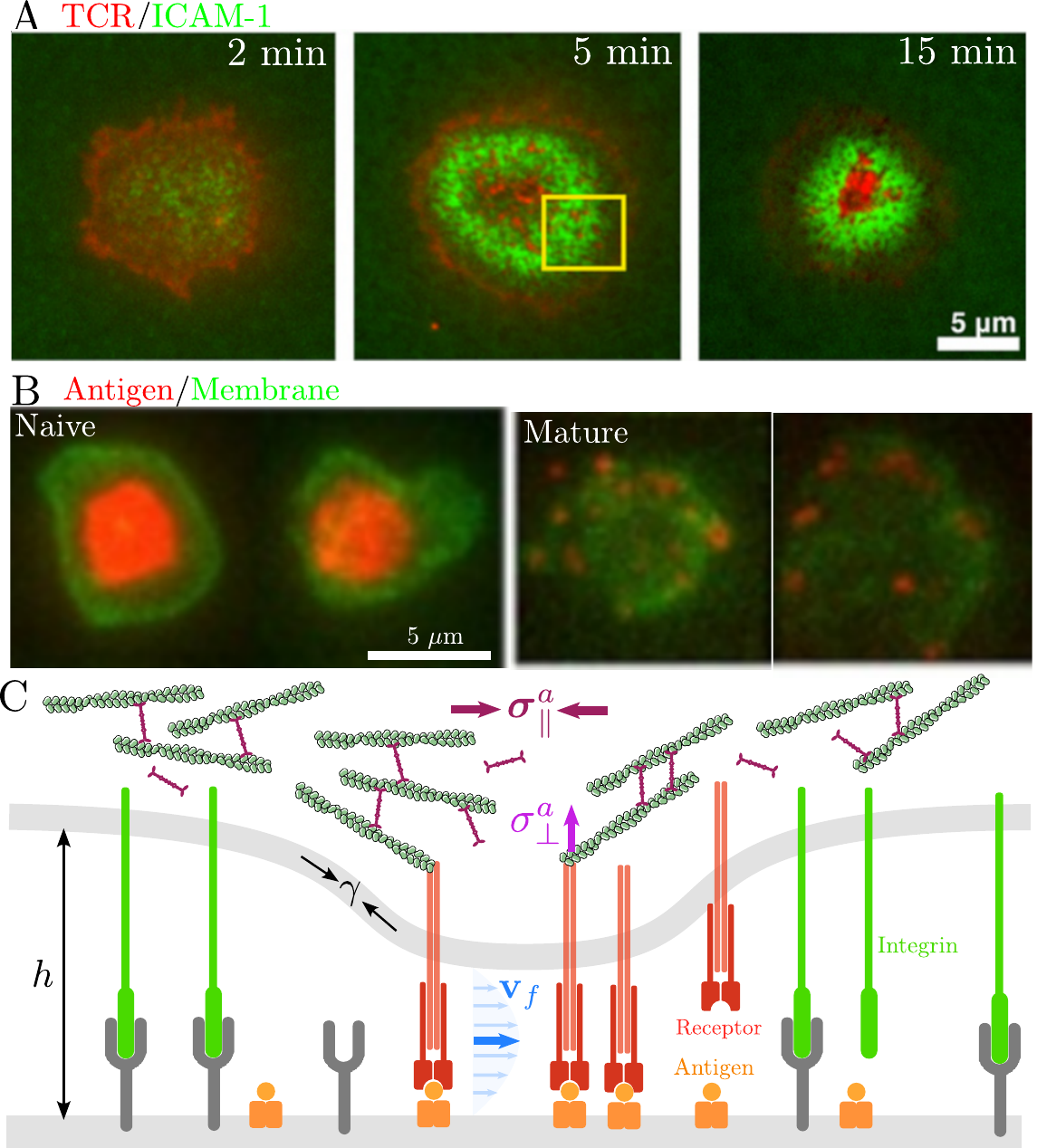}
    \caption{\textbf{The immune synapse is dynamically patterned and mechanically active.} (A) T cells display a centralized
      bull's-eye like pattern of TCRs surrounded by adhesion
      molecules that coarsens over time (adapted from
      Ref.~\cite{hartman2009cluster}). (B) Naive B cells
      display a similar centralized cluster of antigen-bound BCRs, but affinity maturating
      B cells form a distinct pattern of localized
      puncta that enable antigen extraction using cytoskeletal forces (adapted from
      Ref.~\cite{nowosad2016germinal}).  (C) Model schematic showing how receptor-antigen kinetics, integrin exclusion, membrane deformation ($h$),
      cytoskeletal forces ($\vec{\sigma}^{\rm a}_\parallel$, $\sigma^a_{\perp}$), and
      fluid flow (velocity $\b{v}_f$) are coupled.}
    \label{fig1}
\end{figure}

Extensive work has focused on passive mechanisms that drive molecular segregation of receptor and adhesive proteins within cell-cell contact regions \cite{qi2001synaptic,burroughs2002differential,weikl2002pattern,raychaudhuri2003effective,weikl2004pattern,coombs2004equilibrium,figge2009modeling,carlson2015elastohydrodynamics,siokis2017mathematical,knevzevic2018active,prost2024}. Thermodynamic
arguments suggest that differential molecular size coupled with membrane elasticity suffice to generate `bulls-eye' like patterns with short receptor-ligand pairs
concentrated near the center and longer adhesion molecules expelled to
the periphery \cite{coombs2004equilibrium,qi2001synaptic,raychaudhuri2003effective,weikl2004pattern,carlson2015elastohydrodynamics}.
Active transport by cytoskeletal processes was thought to primarily aid in centralizing the pattern~\cite{weikl2004pattern,kaizuka2007mechanisms}, as observed in T cells~\cite{hartman2009cluster,kaizuka2007mechanisms} (see Fig.~\ref{fig1}A).

However, recent studies suggest that T and B cells also use cytoskeletal activity to exert mechanical forces through the immune synapse, affecting synaptic structure \cite{tolar2017cytoskeletal,nowosad2016germinal}, signaling
\cite{basu2017mechanical,huse2017mechanical,upadhyaya2017mechanosensing,rogers2024mechanical}, and affinity discrimination~\cite{brockman2019mechanical,jeffreys2024mechanical,jiang2023immune,jiang2023molecular}. Such mechanisms should be particularly significant in B cells because they use pulling forces to physically extract and internalize antigen \cite{batista2001b,natkanski2013b,kumari2019actomyosin,tolar2017cytoskeletal,jiang2024physical}. Recent experiments also reveal a dramatic influence of cytoskeletal forces on synaptic patterns -- while na{\"i}ve B cells display a centralized cluster with little force application \cite{natkanski2013b}, affinity-maturing B cells exert stronger dynamic forces ($\sim 1-10$~pN) and form distinct multifocal patterns \cite{nowosad2016germinal} (see
Fig.~\ref{fig1}B). These multifocal clusters co-localize with actomyosin and populate the periphery of the contact zone \cite{nowosad2016germinal}, suggesting that active forces can drive pattern formation. A statistical mechanical model has previously captured this patterning transition through a phenomenological account of pattern-dependent force application \cite{knevzevic2018active}. But how do cytoskeletal forces dynamically couple to receptor kinetics to generate diverse synaptic architectures? And how can we understand the functional consequence of the mechanically induced patterns for affinity discrimination?


To describe molecular pattern formation in the immune synapse, we develop a continuum model that couples membrane deformations, receptor kinetics and cytoskeletal activity with elastohydrodynamic flow in the synaptic cleft (see Fig.~\ref{fig1}C). In our model, we go beyond previous approaches \cite{knevzevic2018active,weikl2004pattern,carlson2015elastohydrodynamics,leong2010adhesive} by incorporating both normal and lateral contractile forces derived from the cytoskeleton. Upon numerically solving our model equations, we demonstrate how active cortical flows trigger localized multifocal clusters beyond an activity threshold, while active pulling forces suppress cluster nucleation and enhance their dissolution, resulting in an active coarsening regime. Our results reveal how these dynamic patterns can improve the sensitivity to small affinity differences when active and suggest possible experimental tests of our predictions.


\noindent
\textit{Continuum Model of Synaptic Patterning.} We model the extent of the lubricated gap between an immune cell and an APC by a two-dimensional (2D) circular contact footprint of radius $R$ in which membrane deformations ($h(\b{x},t)$: vertical height) of the immune cell drive squeezing flows
with velocity $\b{v}_f(\b{x},t)$ (see Fig.~\ref{fig1}C). Due to the slender geometry of the gap between apposed cells, we use lubrication
theory \cite{lubrication_rmp_1997} to relate the depth-averaged
velocity $\b{v}_f=-(h^2/12\eta)\vec{\del}p$ ($\eta$ is the fluid
viscosity) to local gradients of fluid pressure $p(\b{x},t)$
(see SM for details).

\begin{figure*}[]
\includegraphics[width=0.99\textwidth]{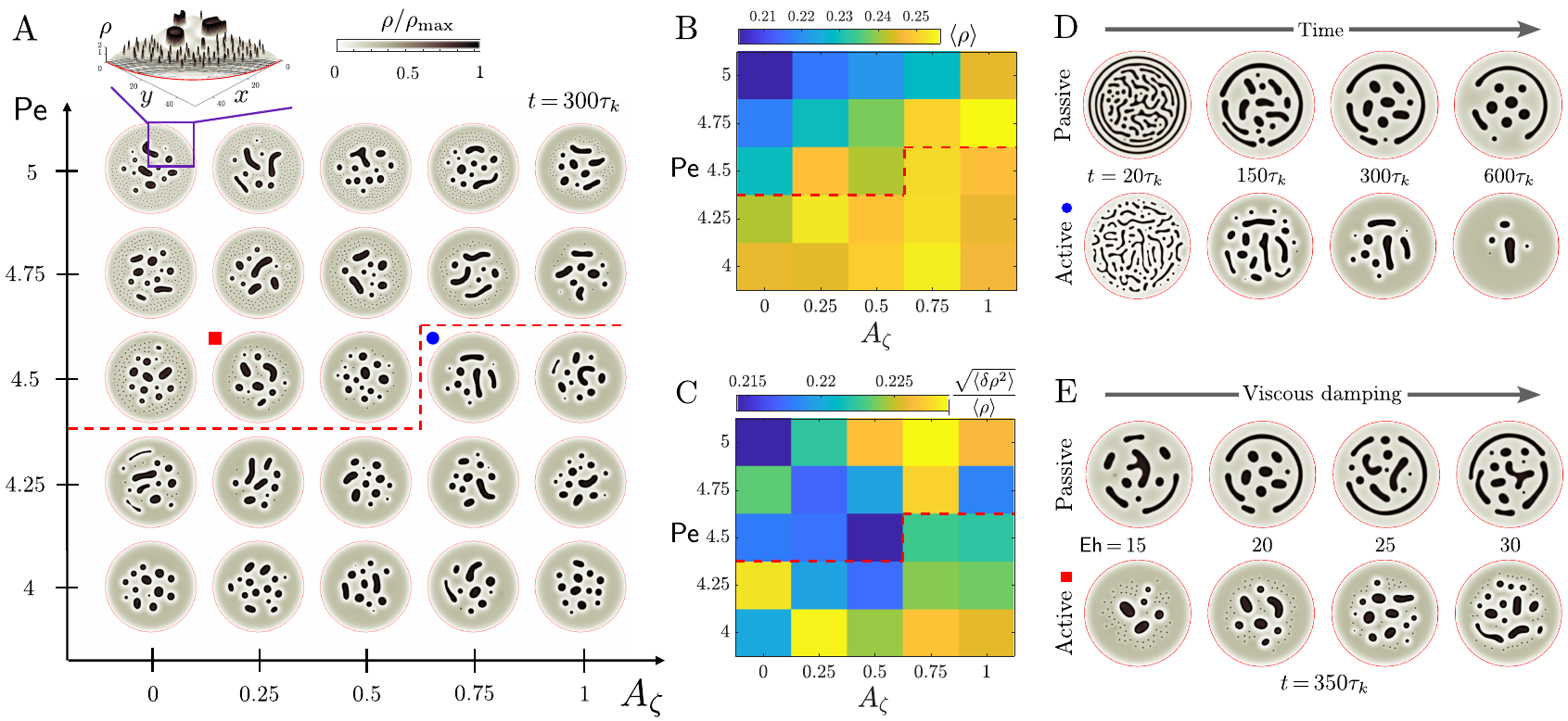}
\caption{{\bf Activity induces transition in synaptic patterns.} 
 (A) State diagram showing representative patterns of normalized bound receptor fraction ($\rho/\rho_{\rm max}$, with $\rho=c_R/c^0_R$) at a fixed time ($t = 300\tau_k$) for varying $\Pe$ and $A_\zeta$. A multifocal pattern with localized puncta emerges when $\mathsf{Pe} \gtrsim  4.25$ (dashed red line), becoming more pronounced for larger $\mathsf{Pe}$. Lower $\Pe$ corresponds to the active coarsening regime. Top inset shows puncta as spikes in $\rho$ near the periphery of the contact zone, with larger domains occupying the center.
  (B-C) Quantifying receptor patterns in (A) through the spatial average $\langle \rho \rangle=(1/\pi R^2)\int\dd\b{x}\;\rho$ (B) and relative fluctuation $\sqrt{\langle \delta\rho^2\rangle}/\langle \rho \rangle$, where $\delta\rho=\rho-\langle\rho\rangle$ (C). (B) Larger $\Pe$ decreases $\langle\rho\rangle$ by creating more, but smaller puncta, while increasing $A_\zeta$ raises $\langle\rho\rangle$ as the background density increases from balanced pushing that overcomes any decrease in $\rho$ due to direct pulling on clusters. (C) The coefficient of variation captures spatial heterogeneity in the pattern, but shows no significant trend with activity. (D) Active coarsening dynamics ($\Pe = 4.5,\,A_\zeta = 0.75$ corresponding to blue circle in A; bottom) proceeds faster than in the passive case ($\Pe = A_\zeta =0$; top) as active contractility breaks apart large domains while pulling forces accelerate dissolution of small clusters and raise the background density of bound receptors. (E) Viscous fluid flow slows pattern formation. Equal time ($t=350\tau_k$) snapshots for the passive ($\Pe=A_\zeta=0$; top) and active ($\Pe=4.5,\,A_\zeta = 0.25$ corresponding to red square in A; bottom) cases show similarity of patterns at small $\mathsf{Eh}$ in short time and large $\mathsf{Eh}$ in long time.}
\label{fig2}
\end{figure*}

Transmembrane proteins associated with short receptor molecules
(length $\ell_R\sim 15~$nm) and long adhesion molecules (length
$\ell_A\sim 40-50~$nm) are assumed to behave as linear springs that
generate forces as they stochastically bind to and unbind from ligands on
the APC. The effective elastic free energy of the membrane
$\mathcal{F}=(1/2)\int\dd\b{x}[\gamma_0(\vec{\del}h)^2+k_Rc_R(h-\ell_R)^2+k_Ac_A(h-\ell_A)^2]$
includes the membrane tension $\gamma_0$ along with the bond stiffness
($k_R,k_A$) and density ($c_R,c_A$) of the bound receptor and adhesion
molecules, respectively. Higher-order bending terms are neglected for
simplicity.

Finally, the actomyosin cortex is described as a thin layer of an
active gel \cite{prost2015active} with a 2D isotropic contractile stress in the plane
$\vec{\sigma}_{\parallel}^a=\alpha m \b{I}$ ($\alpha>0$ is the motor contractility
and $m(\b{x},t)$ is the fraction of phosphorylated actomyosin) that
drives cortical flows $\b{v}_{\rm c}(\b{x},t)$ and lateral transport
of receptor proteins (see SM for details). Vertical active forces enter through the normal component of the active stress
$\sigma^a_{\perp}(\b{x},t)=\zeta [m(\b{x},t)-\langle m\rangle]$ that
exerts pulling forces on the membrane ($\zeta>0$, see SM for
details). These active forces are chosen to be globally (but not locally) balanced by
enforcing a vanishing spatial average ($\langle m\rangle=(1/\pi
R^2)\int\dd\b{x}\;m(\b{x},t)$) so that the entire isolated system is
force-free. Mass conservation and force balance in the vertical and horizontal directions then yield
\begin{equation}
\begin{aligned}
   \partial_th(\b{x},t) & =\vec{\del}\cdot\left(\dfrac{h^3}{12\eta}\vec{\del}p\right)\;,\\
   -p & =-\dfrac{\delta\mathcal{F}}{\delta h}+\sigma^a_{\perp}\;,\\
   -\Gamma\b{v}_{\rm c} + & \vec{\del}\cdot\vec{\sigma}_{\parallel}^a  = 0\;.
\end{aligned}
\label{eq:dhdt}
\end{equation}
$\Gamma$ is an effective frictional drag due to viscous
dissipation in the cortex (see SM for details). Eq.~\ref{eq:dhdt}
emphasizes the key role of elastohydrodynamics, as membrane relaxation
is controlled by spatial pressure gradients rather than occurring locally, as often assumed in ad hoc relaxational models of synaptic
patterning \cite{qi2001synaptic,weikl2004pattern,raychaudhuri2003effective}.
The dynamics of receptor and adhesion proteins obey mass-action kinetics ($i=R,A$) given by 
\begin{align}\label{eq:dcidt}
  \partial_tc_i+\vec{\del}\cdot(c_i\b{v}_i)=\omega^i_{\rm on}(c_i^0-c_i)
  -\omega^i_{\rm off}c_i+D_i\del^2c_i\;,
\end{align}
where cortical flows and membrane deformations generate an advective
flux $\b{v}_i=\b{v}_{\rm c}-(D_i/k_BT)\vec{\del}(\delta\mathcal{F}/\delta
c_i)$ with $D_i\sim 0.5-1~\mu$m$^2$/s a molecular diffusion constant
and $k_{B}T$ the thermal energy. Kinetics are controlled by the
binding and unbinding rates, $\omega_{\rm on}^i(h)$ and $\omega_{\rm
  off}^i(h)$, that depend on the extension/compression of the protein
bonds through local membrane separation $h$. Considerations of
detailed balance then enforce the following relation \cite{dembo1988reaction}
\begin{equation}
    \dfrac{\omega_{\rm on}^i(h)}{\omega_{\rm off}^i}=K^i_{\rm eq}e^{-(k_i/2k_BT)(h-\ell_i)^2}\;,\label{eq:db}
\end{equation}
where $K_{\rm eq}^i$ is the equilibrium binding affinity (association
constant at zero load) of the bound protein complex. Here, we choose a
constant unbinding rate and leave the study of force-dependent
unbinding (e.g., catch or slip behavior) to future study. For
computational ease and to reduce the number of parameters, we adopt a
further simplification by assuming that the kinetics of the adhesion molecules
are rapidly equilibrated \cite{raychaudhuri2003effective} so that
$c_A(h)\approx c_A^0\omega_{\rm on}^A(h)/[\omega_{\rm
    on}^A(h)+\omega_{\rm off}^A]$ is locally specified by the height of the membrane. This approximation retains the activity-mediated instabilities
in the receptor dynamics and is sufficient for our purposes (see SM
for further justification), although a complete analysis of the full
model is left as an open problem.

To close our equations, we prescribe how motor activity depends
on the molecular organization of the synapse. Consistent with
experimental observations of actomyosin enrichment near
antigen-bound B cell receptor clusters \cite{nowosad2016germinal}, we invoke a feedback
law that relates the local phosphorylated actomyosin  fraction to the bound receptor concentration via a Michaelis-Menten-like response
\begin{equation}
    m = m_0\left(\dfrac{1+\chi c_R}{1+m_0\chi c_R}\right)\;,\label{eq:m}
\end{equation}
controlled by a basal actomyosin fraction $m_0$ (present even when
$c_R=0$) and activation sensitivity $\chi>0$. Eq.~\ref{eq:m}
assumes fast actomyosin recruitment and neglects
delays in the build-up of active stresses. Finally, to complete our model, we impose boundary conditions at the edge of the circular contact zone that pin the
membrane height ($h(R)=h_0$), allow free drainage of fluid ($p(R)=0$),
and set a vanishing diffusive flux of bound receptors
$\hat{\vec{\nu}}\cdot\vec{\del}c_R|_{R}=0$ ($\hat{\vec{\nu}}$ is the outward
unit normal).


\noindent
\textit{Scaling and Nondimensional Parameters.} A simple scaling analysis reveals the key features of our model
(Eqs.~\ref{eq:dhdt}-\ref{eq:m}). Without activity
($\alpha=\zeta=0$), bound receptors phase separate from adhesion molecules, forming domains \cite{carlson2015elastohydrodynamics} of a
characteristic size $L_{\rm c}\sim\sqrt{\gamma_0/(k_Rc_R^0)}$ that balances membrane tension and receptor stretching. For an
average membrane separation $h_0$, the receptor density equilibrates
on a fast time scale $\tau_k=[\omega_{\rm
    on}^R(h_0)+\omega_{\rm off}^R]^{-1}$, while patterning dynamics on a length $L$ occur on the
elastohydrodynamic time scale $\tau_h\sim
(\eta/k_Rh_0c_R^0)(L/h_0)^2$. The relative importance of kinetics
($\tau_k$) to fluid flow ($\tau_h$) is measured by an
\textit{elastohydrodynamic number}: $\mathsf{Eh}=(12\eta/k_Rc_R^0\rho_0h_0\tau_k)(L_{\rm c}/h_0)^2$,
%
where $\rho_0=\omega^R_{\rm on}(h_0)\tau_k$ is the bound receptor
fraction. Using typical values $\eta\sim 10^{-3}~$Pa~s, $\gamma_0\sim
0.03~$mN/m, $k_R\sim 0.01-0.3~$pN/nm, $c_R^0\sim 200~\mu$m$^{-2}$,
$h_0\sim 35~$nm, and $\tau_k\sim 0.1-10~$s \cite{qi2001synaptic,carlson2015elastohydrodynamics,weikl2004pattern} gives $L_{\rm c}\sim
50-100~$nm and a wide range of values $\Eh\sim 10^{-3}-10$.

\begin{figure*}[]
\centering
  \includegraphics[width=0.99\textwidth]{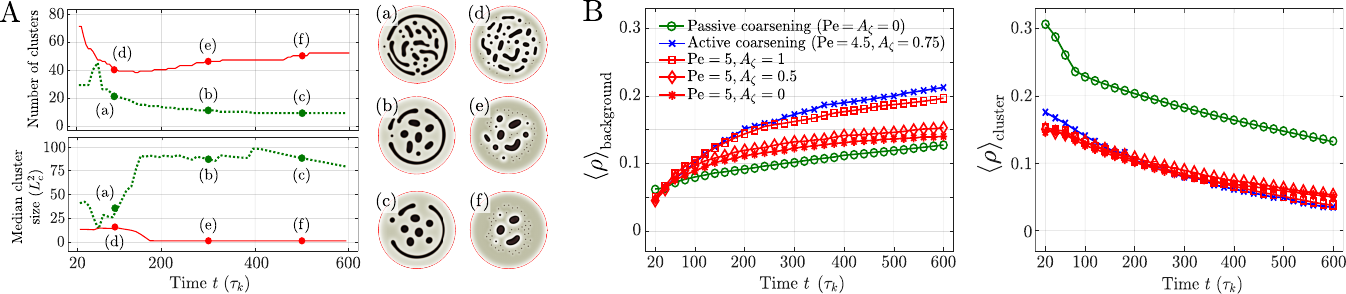} 
  \caption{
{\bf Dynamics of active patterns.} (A) Time evolution of cluster number and size distinguishes active multifocal patterning from passive coarsening. Left panels show temporal trajectories of cluster number and median size in passive (green dashed) and active (red solid; $\Pe=4.5,A_\zeta=0.25$) scenarios, with sample patterns at labeled time points for passive (a-c) and active (d-f) cases shown on the right. In the active case, the median cluster size at long time plateaus at the typical size $\sim  1.01L_c^2$ of the puncta. Our simulations resolve individual puncta using a mesh size of at most one third of the puncta radius. (B)
Time evolution of the bound fraction of receptors in the background (left) versus within clusters (right), per unit domain area for: (i) passive coarsening ($\Pe = A_\zeta = 0$; green), (ii) active coarsening ($\Pe = 4.5$, $A_\zeta = 0.75$; blue), and (iii) active multifocal patterns ($\Pe = 5$ and varying $A_\zeta$; red). Note, $\langle\rho\rangle_{\Omega}(t)=(1/\pi R^2)\int_{\rm \Omega}\dd\b{x}\;\rho(\b{x},t)$, where $\Omega=\{\rm cluster,\,background\}$, so the spatial average $\langle\rho\rangle=\langle\rho\rangle_{\rm cluster}+\langle\rho\rangle_{\rm background}$. See Fig.~S3 for the time trace of the total bound fraction $\langle \rho \rangle$.
}
\label{fig3}
\end{figure*}

Activity introduces additional scales to the problem. Spatial
gradients of active stresses acting over a length $L$ drive horizontal flows
$|\b{v}_{\rm c}|\sim \alpha m_0\chi c_R^0/(\Gamma L)$ that enhance receptor
clustering, while vertical active forces
$\sim \sigma^a_{\perp}/c_R\sim \zeta m_0\chi$ stretch
receptors and suppress antigen binding. Combining active and passive
processes in both horizontal and vertical directions defines two key
parameters -- a P{\'e}clet ($\Pe$) and activity ($A_\zeta$) number
 defined by
\begin{equation}
  \Pe=\alpha \dfrac{m_0(1-m_0)\chi c_R^0\rho_0}{\Gamma D_R}\;,
  \quad A_\zeta=\zeta\dfrac{m_0(1-m_0)\chi}{k_Rh_0}\;,
\end{equation}
which quantify the relative importance of advection compared to
diffusion and active pulling compared to bond elasticity,
respectively. Assuming $\chi\sim 1$ and using $\Gamma\sim
10^2-10^3~$Pa~s/$\mu$m$^2$ for cortical friction, $\alpha m_0\sim
0.1-10~$kPa for contractility \cite{salbreux2012actin,joanny2009active}, and $\zeta m_0\sim 10-50~$Pa for pulling stresses \cite{natkanski2013b,upadhyaya2017mechanosensing}, we
find the relevant ranges $\Pe\sim 0.1-10$ and $A_\zeta\sim 10^{-2}-3$
to explore. 


\noindent
\textit{Results.}
We numerically solve Eqs.~\ref{eq:dhdt}-\ref{eq:m} using a custom
finite element code implemented using FEniCS \cite{BarattaEtal2023} (see SM for
details). We choose the kinetic time ($\tau_k$) and typical domain size ($L_c$) to nondimensionalize time and length. To investigate the role of cytoskeletal activity, we fix the parameters of the passive model using experimental estimates (see SM for details) and construct a state diagram of synaptic patterns for varying $\Pe$ and
$A_\zeta$, quantified using the spatial mean and variance of receptor density profiles (Fig.~\ref{fig2}A-C). For
$\Pe=A_\zeta=0$, we recover the passive phase separation of receptor proteins and adhesion molecules (Fig.~\ref{fig2}D top row; Movie 1) \cite{carlson2015elastohydrodynamics}.
%
%

Increasing vertical pulling $A_\zeta$ stretches receptor molecules, reducing the miscibility gap (see SM for details) and accelerating both coarsening and dissolution of patterns -- a form of `active coarsening' (Fig.~\ref{fig2}D bottom row; Movie 2). By contrast, increasing contractility $\Pe$ promotes smaller, denser clusters.  For $\Pe \geq 4.25$ with zero pulling force ($A_\zeta = 0)$, we obtain a contractile instability that creates long-lived, localized, high-density puncta at the domain boundary (Fig.~\ref{fig2}A and inset; Movie 3), reminiscent of multifocal patterns seen in mature B cells (Fig.~\ref{fig1}B). While weak pulling forces (small $A_\zeta$) preserve the multifocal pattern, stronger pulling (larger $A_\zeta$) speeds up cluster dissolution and suppresses puncta formation, eventually eliminating the remaining domains of bound receptors as the membrane fully delaminates (see red transition boundary in Fig.~\ref{fig2}A). 
This transition is not captured by linear stability analysis (see SM for details); instead, puncta emerge nonlinearly, forming localized structures similar to those recently reported in active fluids \cite{barberi2023localized}. Their even spacing suggests an effective repulsion between puncta that is mediated by nonlinear density gradients and membrane deformations.

Viscous fluid flow causes slower dynamics upon increasing $\Eh$, in both passive ($\Pe=A_\zeta=0$) and active ($\Pe=4.5$, $A_\zeta=0.25$) patterns; see Fig.~\ref{fig2}E.
However, distinct spatial structures emerge, which we quantify using the number and typical size of clusters over time  (Fig.~\ref{fig3}A). In the passive limit ($\Pe=A_\zeta=0$), large domains of bound receptors grow, round up, and dissolve slowly, but smaller clusters shrink and annihilate, continuously reducing the number of clusters (Fig.~\ref{fig3}A, top, green  line; Movie 1). The disappearance of small clusters leads to erratic increases of the median cluster size, e.g., in $t \in [200,400]\,\tau_k$ (Fig.~\ref{fig3}A, bottom, green line), while slow dissolution of large domains causes the median size to steadily decrease, as is most apparent for $t \in [400,600]\,\tau_k$. In the active multifocal clustering regime ($\Pe=4.5$, $A_\zeta=0.25$), we instead observe that the number of clusters increases as the median cluster size drops to a steady value $\sim L_c^2$ (Fig.~\ref{fig3}A, red lines). This reflects the rising abundance of mutually repelling puncta as large clusters shrink and smaller clusters dissolve. 
  The stable assembly of localized puncta persists for extended periods of time and dominates the pattern for $t \geq 200\,\tau_k$ (Fig.~\ref{fig3}A, pattern (f) and Movie 3). 

To globally quantify synaptic pattern dynamics, we measure the time evolution of the spatially averaged bound receptor fraction ($\rho=c_R/c_R^0$), distinguishing the contributions from the clusters ($\langle \rho \rangle_{\rm cluster}$) and the background ($\langle \rho \rangle_{\rm background}$); see Fig.~\ref{fig3}B.  In the passive case, after an initial transient, cluster dissolution and coarsening cause $\langle \rho \rangle_{\rm cluster}$ to decrease and $\langle \rho \rangle_{\rm background}$ to slowly increase with time. Activity retains these temporal trends, only changing their amplitudes. Strong activity suppresses $\langle \rho \rangle_{\rm cluster}$ relative to its passive counterpart, as smaller clusters (rather than large domains) form. But once formed, the temporal dynamics is insensitive to changes in activity, in both the active coarsening (Fig.~\ref{fig3}B, blue line) and multifocal patterning (Fig.~\ref{fig3}B, red lines) regimes.

However, $\langle \rho \rangle_{\rm background}$ grows with time and is enhanced by increasing $A_\zeta$ (Fig.~\ref{fig3}B), buffering the overall decay in $\langle \rho \rangle$ (see Fig.~S3). This is consistent with an effective free energy construction (see SM for details) showing that active pulling raises the coexisting density outside clusters. Indeed, active coarsening results in a higher $\langle \rho \rangle_{\rm background}$ than multifocal patterning at $\Pe=5$ (Fig.~\ref{fig3}B), due to an accelerated dissolution of transient clusters. Active forces thus provide a buffering mechanism, tuning the bound receptor fraction in solution without altering that within clusters.

\begin{figure}[]
\includegraphics[width=0.38\textwidth]{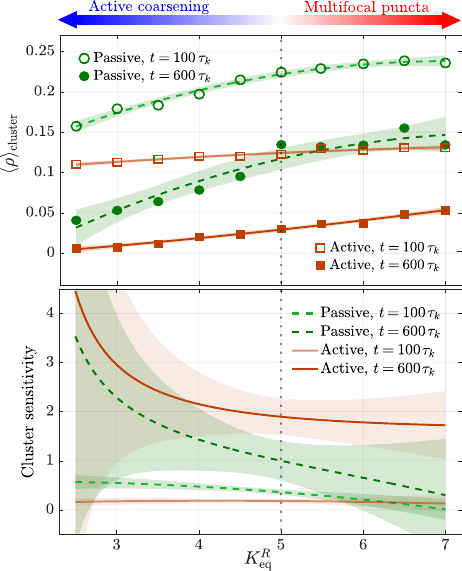}
\caption{{\bf Active patterning allows sensitive discrimination over a wide affinity range.}
Top: The bound fraction of receptors in clusters $\langle \rho \rangle_{\rm cluster}$ as a function of  $\KR$ in passive (green) and active (red) systems at $t=100 \,\tau_k$ (open symbols)
 and $t=600 \, \tau_k$ (filled symbols). Lines are quadratic fits with $95\%$ confidence interval shown by the shading. 
 Bottom: Logarithmic cluster sensitivity computed as $\partial\ln\langle \rho \rangle_{\rm cluster}/\partial\ln\KR$ obtained from the fitted curves in the top panel. In the active case, $\Pe=4.5$ and $A_\zeta=0.25$ are fixed; as $\KR$ increases, the system transitions from active coarsening to multifocal clustering around $\KR\simeq 5$.}
\label{fig4}
\end{figure}

Finally, we ask how these different patterns can impact
affinity discrimination. We consider $\langle \rho \rangle_{\rm cluster}$ as a proxy for signaling strength since experiments indicate that sufficient clustering precedes signaling~\cite{GROVES2011, ketchum2014ligand}. 
Discrimination accuracy then boils down to the sensitivity of $\langle \rho \rangle_{\rm cluster}$ to small differences in receptor affinity $\KR$. Fixing $\Pe=4.5,A_\zeta=0.25$ and increasing $\KR$, the system transitions from active coarsening to multifocal patterns at $\KR\simeq 5$, and we find $\langle \rho \rangle_{\rm cluster}$ is consistently higher in the passive case (green) than in the active case (red), see  Fig.~\ref{fig4}B (top). This is consistent with experimental observations of germinal center B cells that form puncta and apply considerable forces, but gather a smaller amount of antigen compared to na{\"i}ve B cells that exert little force~\cite{nowosad2016germinal}. While the cluster component $\langle \rho \rangle_{\rm cluster}$ decreases with time (open vs.~filled symbols) in both active and passive cases, the total bound receptor fraction $\langle \rho \rangle$ shows a pattern-dependent trend: it increases with time for $\KR \ge 5$, where multifocal patterns emerge, but decreases with time for $\KR \le 4.5$, when active coarsening occurs instead (see Fig.~S4 and~S5).
  
The logarithmic sensitivity of $\langle \rho \rangle_{\rm cluster}$ to $\KR$ (Fig.~\ref{fig4}, bottom; see Fig.~S4 for sensitivity of $\langle\rho\rangle$) reveals an affinity-dependent speed-sensitivity trade-off. At early times ($t=100\,\tau_k$) and low $\KR$, the passive system exhibits a moderately higher sensitivity than the active system; this trend reverses at larger $\KR$. At late times ($t=600\,\tau_k$), the active system instead shows a significantly higher sensitivity for all $\KR$ studied; the excess in sensitivity over that of the passive system is more pronounced at higher affinity. This suggests long-lived puncta that form for $\KR \ge 5$ are likely responsible for the enhanced sensitivity. This hypothesis can be tested by reducing receptor diffusion (e.g., through membrane cholesterol depletion) thus raising $\Pe$ while keeping $A_\zeta$ intact. Higher $\Pe$ (stronger contractility relative to pulling) is expected to speed up multifocal clustering, resulting in a faster gain in sensitivity. 

\noindent
\textit{Conclusions.} By coupling both horizontal (contractile) and vertical (pulling) active forces with molecular kinetics and elastohydrodynamics of membrane adhesion, we have shown how a minimal description can capture the emergence of localized multifocal patterns (for large $\Pe$) and transition to active coarsening dynamics (for large $A_\zeta$), consistent with \textit{ex vivo} experiments. The slender geometry of the synaptic cleft lets squeezing flows make membrane relaxation non-local and sensitive to boundary conditions, while membrane tension generates an effective repulsion between short receptor-antigen complexes and long adhesion molecules. We identify a new class of activity-induced localized states and demonstrate that feedback between molecular organization and active force exertion is key to stabilizing multifocal patterns~\cite{knevzevic2018active}.

Our results show that synaptic patterning influences affinity discrimination by adaptively controlling speed, accuracy, and dynamic range. Nonequilibrium activity enables a greater sensitivity that persists as puncta form above a threshold affinity, unlike passive coarsening in which discrimination degrades with affinity. Although multifocal recognition is known to optimize information acquisition~\cite{jiang2024physical}, competition between active forces yields new behavior: it can tunably partition between localized clusters and dispersed receptors. This flexible partitioning can balance the dynamic range with sensitivity, as in chemotaxis~\cite{bray1998receptor}, suggesting a physical mechanism for rapid adaptation of immune responses to external conditions such as membrane state and antigen properties. 

\noindent{\it Acknowledgments.} We thank L.~Mahadevan for insightful discussions in the early phase of this work.
This research was supported in part by NSF grant PHY-2309135 to the Kavli Institute for Theoretical Physics (KITP), and by grant 815891 to the Simons Foundation, and through computational resources and services provided by Advanced Research Computing at the University of Michigan, Ann Arbor. SW is grateful for funding support from the National Science Foundation (NSF) Grant MCB-2225947 and an NSF CAREER Award PHY-2146581. 
\end{document}